\documentclass[aps,twocolumn,amsmath,amssymb]{revtex4}
\usepackage{graphicx}
\usepackage{color}
\usepackage{amsfonts}

\def\g5{\gamma_{5}}
\def\ga{\gamma}

\def\la{\lambda}

\def\be{\begin{eqnarray}}
\def\ed{\end{eqnarray}}

\def\non{\nonumber}

\def\la{\langle}
\def\ra{\rangle}

\begin{document}
\title{  Comment on
``New Physics Contributions to the \\ Lifetime Difference in
$D^0-\bar D^{0}$ mixing''}
\author{  Chuan-Hung Chen$^{1,2}$,
Chao-Qiang Geng$^{3,4}$
and
 Soo-Hyeon Nam$^{5}$
 }

\affiliation{ $^{1}$Department of Physics, National Cheng-Kung
University, Tainan 701, Taiwan \\
$^{2}$National Center for Theoretical Sciences, Taiwan\\
$^{3}$Department of Physics, National Tsing-Hua University, Hsinchu
300, Taiwan  \\
$^{4}$Theory Group, TRIUMF, 4004 Wesbrook Mall, Vancouver, B.C V6T
2A3, Canada \\
$^{5}$Department of Physics, National Central University, Chung-Li
32054, Taiwan }

\maketitle

\begin{figure}[htbp]
\includegraphics*[width=2.0 in]{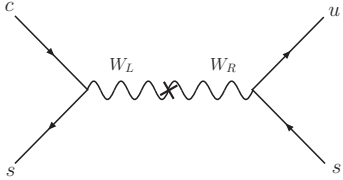}
\caption{Feynman diagram for the left-right mixing.}
 \label{fig:lr}
\end{figure}

The precision
measurement on the oscillation of a neutral particle and its
antiparticle is very important to probe the footprints of new
physics (NP). 
Although so far no clear signal tells us the existence of
NP,
Golowich, Pakvasa and Petrov \cite{GPP_PRL98}
have recently shown
that the NP effects
could have significant contributions
to the lifetime difference in the $D^{0}-\bar D^{0}$ mixing,
with the resultant form
 \be
 y\simeq \sum_{n} \frac{\rho_n}{\Gamma_D} A^{(SM)}_n \bar A^{(SM)}_n
 + 2 \sum_{n} \frac{\rho_n}{\Gamma_D} A^{NP}_n \bar A^{(SM)}_n\,,
 \label{eq:y}
 \ed
where
$y\equiv \Delta \Gamma_D/2\Gamma_D$,
$\Gamma_D\, (\Delta\Gamma_D)$ is
the decay rate (rate difference),
$\rho_n$ corresponds to the phase space of the
charmless intermediate
state n and $A^{(SM)} [A^{(NP)}]$ denotes the decay amplitude for
$D^{0}\to n$
dictated by the standard model (SM) [NP] contributions. In their
Letter,
various NP models, including
vectorlike, SUSY without (with) R-parity, left-right (LR) and
multi-Higgs models, on the Eq.~(\ref{eq:y}) are analyzed. Their
results display that except the SUSY models without R-parity,  $y$
is typically less than $10^{-4}$.
However, we find that in the nonmanifest LR model, an important LR
mixing effect, illustrated in
Fig.~\ref{fig:lr},
was not considered in Ref. \cite{GPP_PRL98}.

Two charged gauge bosons $W_{L,R}$ in the LR model are mixed and
described by $W_1=\cos \zeta W_L + \sin\zeta W_R$ and
$W_2=-\sin\zeta W_L + \cos\zeta W_R$ \cite{PDG06}, where $W_1$ is
the observed boson and $\zeta$ is the mixing angle between $W_L$ and
$W_R$.
 Interestingly, $\zeta$ of
$O(10^{-2})$ is still allowed by current experimental data
\cite{PDG06,Nam_PRD68}. Using the small mixing angle, the charged
current interaction associated with LR mixing is given by
  \be
 {\cal L}_{LR}&=& \frac{g_R}{\sqrt{2}} V^{(R)}_{ab} \zeta
  \bar u_b \ga_{\mu}  P_R  d_a  W^{\mu}_{1}\,,
 \ed
where $a(b)$ is the flavor index. $g_R$ and $V^{(R)}$ are the
right-handed gauge coupling and flavor mixing matrix, respectively.
In the nonmanifest LR model, since the patten for $V^{(R)}$ could be
$V^{(R)}_{cs}\sim V^{(R)}_{ud}\sim 0$ and $V^{(R)}_{us}\approx
V^{(R)}_{cd}\approx O(1)$ \cite{LS_PRD40}, it is clear that
Fig.~\ref{fig:lr} gives the dominant contribution. In terms of the
leading results of the operator product expansion,
only
${\cal O}^{ijk\ell}_4= \bar u_{k} \Gamma_{\mu}
\slash{\!\!\!{p}}_{c}
\bar \Gamma_2 c_j\, \bar u_{\ell} \bar \Gamma_1 \Gamma^{\mu} c_i$ in
Eq.~(7) of Ref.~\cite{GPP_PRL98} has the contribution, where $i,\,
j,\, k,\, \ell$ denote the color indices, $\Gamma_{\mu}=\ga_{\mu}
P_L$, $\bar\Gamma_2=\ga_{\nu} P_L$ and $\bar \Gamma_1=\ga^{\nu}
P_R$. Consequently, the effect for the lifetime difference is found
to be
 \be
y_{LR}&=& {\cal C}_{LR} V^{(L)}_{cs} V^{(R)^*}_{us} \left[ K_{2}
\langle Q' \rangle + K_{1} \langle \tilde Q' \rangle \right]\,,
 \ed
 where
 $K_1=C_1 \bar C_1
N_c + C_1 \bar C_2 + \bar C_1 C_2$ with
$N_c=3$, $K_2=C_2 \bar C_2$,  $\langle Q'
\rangle = \langle \bar D^{0} | \bar u_i \gamma_{\mu} P_{L} c_i \bar
u_j \gamma^{\mu} P_{R} c_j | D^0\rangle$, $\langle \tilde Q'
\rangle = \langle \bar D^{0} | \bar u_i \gamma_{\mu} P_{L} c_j \bar
u_j \gamma^{\mu} P_{R} c_i | D^0\rangle$
and
 \be
 {\cal C}_{LR}&=&\frac{G^2_{F}}{2 \pi m_D \Gamma_D}  \xi_g \lambda m^2_c
 \sqrt{x_s}\,, \label{eq:cLR}
 \ed
with
$x_s=m^2_s/m^2_c$, $\xi_g=\zeta g_R/g_L$ and $\lambda$ being the Wolfsenstein's parameter.
Here, $\bar C_{1(2)}=C_{1(2)}$.
 From Eq.~(\ref{eq:cLR}),
we see clearly that the LR mixing contribution is at least
$1/\sqrt{x_s}=m_c/m_s$ larger than that presented in
Ref.~\cite{GPP_PRL98}.

With $G_{F}=1.166\times 10^{-5}$ GeV$^{-2}$, $m_s=0.12$ GeV,
$m_c=1.4$ GeV, $m_D=1.86$ GeV, $f_D=0.23$ GeV, $\xi_{g}< 0.033$,
$C_1\approx -0.55$ and $C_2\approx 1.3$,  we get the maximum value
of
 \be
  y_{LR}\approx  1.4 \times 10^{-3}\,.
  \label{yLR}
 \ed

Recently, BABAR \cite{babar_D} and BELLE \cite{belle_D}
collaborations have reported the evidence for the $D-\bar D$ mixing
with
the combined $68\%$ C.L. results of
\be
\label{D12}
x&\equiv &{\Delta m_D\over \Gamma_D}= (5.5\pm 2.2)\times 10^{-3}\,,
\non\\
y&=& (5.4\pm2.0)\times 10^{-3}\,.
\ed
%
Clearly, our result in Eq. (\ref{yLR}) based on the LR mixing mechanism in the nonmanifest LR model fits well with the data in Eq. (\ref{D12}).



\end{document}